\documentclass[a4paper,11pt]{article}
\pdfoutput=1 
\usepackage{jheppub}  
\usepackage{graphicx,color}
\usepackage{amsmath}
\usepackage{autobreak}
\usepackage{array}
\newcolumntype{P}[1]{>{\centering\arraybackslash}p{#1}}
\allowdisplaybreaks
\newcommand{\ben}{\begin{enumerate}}
\newcommand{\een}{\end{enumerate}}
\newcommand{\beq}{\begin{equation}}
\newcommand{\eeq}{\end{equation}}
\newcommand{\bal}{\begin{align}}
\newcommand{\eal}{\end{align}}

\def\gSVN#1{{\Delta_{sv}^{(#1)}}}
\newcommand{\Ca}{C_A}
\newcommand{\Cf}{C_F}
\newcommand{\nf}{n_f}

\newcommand{\Lqr}{L_{qr}}
\newcommand{\Lfr}{L_{fr}}
\newcommand{\Lqrfr}{L_{qr}L_{fr}}
\newcommand{\Lqrtfr}{L_{qr}^{2}L_{fr}}
\newcommand{\Lqrfrt}{L_{qr}L_{fr}^{2}}
\newcommand{\LNb}{\bar{L}}
\def\z#1{{\zeta_{#1}}}
\def\g0#1bbH{{g_{b,0}^{(#1)}}}
\def\gNB#1{{g_{#1}}}

\newcommand{\iW}{\lambda^{-1}}
\def\LogmW1{{{\ln (1-\lambda)}}}
\def\w{{\lambda}}
\def\WbimW{{\frac{\lambda}{(1-\lambda)}}}
\def\LogomWtIMW{{\frac{\ln(1-\lambda)}{(1-\lambda)}}}
\def\LogomWtIMWt{{\frac{\ln (1-\lambda)}{(1-\lambda)^2}}}
\def\LogtmWtIMWt{{\frac{\ln (1-\lambda)^2}{(1-\lambda)^2}}}
\def\LogttmWtIMWt{{\frac{\ln (1-\lambda)^3}{(1-\lambda)^2}}}

\def\LogtmWtIMW{{\frac{\ln (1-\lambda)^2}{(1-\lambda)}}}
\def\WttmWtimWt{{\frac{\lambda(2-\lambda)}{(1-\lambda)^2}}}
\def\WtbimWt{{\frac{\lambda^2}{(1-\lambda)^2}}}

\def\AAo{{\frac{A_1^{(b)}}{\beta_{0}}}}
\def\AAt{{\frac{A_2^{(b)}}{\beta_{0}^2}}}

\def\DDo{{\frac{D_1^{(b)}}{\beta_{0}}}}

\def\btzAIII{{\frac{A_3^{(b)}}{\beta_0^2}}}
\def\btzAII{{\frac{A_2^{(b)}}{\beta_0}}}
\def\btzAI{{A_1^{(b)}}}

\def\btzDII{{\frac{D_2^{(b)}}{\beta_0}}}
\def\btzDI{{D_1^{(b)}}}

\def\btztAIV{{\frac{A_4^{(b)}}{\beta_0^2}}}
\def\btztAIII{{\frac{A_3^{(b)}}{\beta_0}}}
\def\btztAII{{A_2^{(b)}}}
\def\btztAI{{\beta_0 A_1^{(b)}}}

\def\btztDIII{{\frac{D_3^{(b)}}{\beta_0}}}
\def\btztDII{{D_2^{(b)}}}
\def\btztDI{{\beta_0 D_1^{(b)}}}

\def\bobt{{\frac{\beta_1 \beta_2}{\beta_0^5}}}

\newcommand{\btoo}{{\bigg(\frac{\beta_{2}}{\beta_0^{3}}\bigg)}}
\newcommand{\bthr}{{\bigg(\frac{\beta_{3}}{\beta_0^{4}}\bigg)}}

\title{Resummed prediction for Higgs boson production through $b\bar{b}$ annihilation at N$^3$LL}
\author[a]{Ajjath A H,}
\author[a]{Amlan Chakraborty,}
\author[b]{Goutam Das,}
\author[a]{Pooja Mukherjee,}
\author[a]{V. Ravindran}
\affiliation[a]{The Institute of Mathematical Sciences, HBNI, IV Cross Road, Taramani, Chennai 600113, India}
\affiliation[b]{Theory Group, Deutsches Elektronen-Synchrotron (DESY), Notkestrasse 85, D-22607 Hamburg, Germany}
\emailAdd{ajjathah@imsc.res.in}
\emailAdd{amlanchak@imsc.res.in}
\emailAdd{goutam.das@desy.de}
\emailAdd{poojamukherjee@imsc.res.in}
\emailAdd{ravindra@imsc.res.in}
\abstract{
We present an accurate theoretical prediction for the production of Higgs boson through bottom quark
annihilation at the LHC up to next-to-next-to-next-to leading order (N$^3$LO) 
plus next-to-next-to-next-to-leading logarithmic 
(N$^3$LL) accuracy.  We determine the third order perturbative Quantum Chromodynamics (QCD) correction 
to the process dependent constant in the resummed
expression using the three loop bottom quark form factor and third order quark soft distribution function. 
Thanks to the recent computation of N$^3$LO corrections to this
production cross-section from all the partonic channels, an accurate matching can be obtained 
for a consistent predictions at N$^3$LO+N$^3$LL accuracy in QCD. 
We have studied in detail the impact of resummed threshold contributions to inclusive cross-sections 
at various centre-of-mass energies and also discussed their sensitivity to renormalization and factorization
scales at  next-to-next-to leading order (NNLO) matched with next-to-next-to leading logarithm (NNLL). At  N$^3$LO+N$^3$LL, we predict the cross-section for different centre-of-mass energies using the recently available results in  \cite{Duhr:2019kwi} as well as study the renormalization scale dependence at the same order.
}

\begin{document} 
\preprint{IMSC/2019/05/04, DESY 19-076}
\maketitle
\section{Introduction} \label{intro}
Discovery of the Higgs boson in $2012$ is one of the biggest achievements of ALTAS \cite{Aad:2012tfa} and
CMS \cite{Chatrchyan:2012xdj} collaborations at the Large Hadron Collider (LHC).
It is a milestone in the success of the Standard Model (SM).  Having understood the generation of 
mass for the elementary particles through spontaneous symmetry breaking, it is important to 
understand the properties \cite{Heinemeyer:2013tqa} of the Higgs boson such as spin, CP properties, 
self coupling and the couplings to the SM fermions and vector bosons.  
In the SM, Higgs boson is spin-0 and CP-even and it couples to SM fermions 
through Yukawa coupling.   
There exists several beyond the SM (BSM) scenarios which allow 
non SM spin-0 or spin-2 bosons to couple to SM particles.  In addition, there exists  
CP mixing in the extended Higgs sectors. All these scenarios demonstrate 
distinct observable effects which can be studied at the LHC thereby constraining various BSM physics.  

In recent years, efforts to understand the shape of the Higgs potential by measuring the Higgs self 
coupling in di-Higgs boson production is underway.  It has important implications for the 
hierarchy problem, the vacuum metastability,  the electroweak phase transition and baryogenesis.
The measurement of the di-Higgs boson production at the LHC is plagued by the tiny cross-section
\cite{deFlorian:2016spz},  however, the high luminosity option can help.  

Yukawa couplings of the Higgs boson to the quarks and leptons of the SM are free parameters and hence, understanding their pattern is of great importance.    
As the mass of the Higgs boson is close to the electroweak scale, these couplings are highly sensitive to scales
of new physics and measuring them precisely can probe various high scale physics scenarios.  Both
ATLAS \cite{ATLAS-CONF-2019-005} and CMS \cite{Sirunyan:2018koj} collaborations have made 
dedicated efforts to measure 
them in various decay channels of the Higgs boson.   
Yukawa coupling of bottom quarks to the Higgs boson is one of the most sought parameter. 
However, measuring this coupling in the dominant decay channel of the Higgs boson to a pair of bottom quarks 
is a challenging task.    
Associated production of Higgs boson with vector bosons or with top quarks, its subsequent decay to bottom quarks 
is the promising one  while there are also other proposals \cite{Englert:2015dlp}.

While the Higgs boson dominantly decays to bottom quarks in SM, its production
from bottom quark annihilation is much smaller than the gluon initiated subprocess.  
However,  as the precise measurement of the Higgs cross-sections is underway at the LHC, inclusion of 
bottom quark initiated channels in the theoretical predictions is unavoidable.  
Unlike in the SM, bottom quarks in the Minimal Supersymmetric SM (MSSM) \cite{Gunion:1992hs}
couple to neutral Higgs boson with the coupling proportional to $1/\cos\beta$ which in some
parameter region can increase the production rate.  Here, the angle $\beta$ is related to the ratio, denoted
by $\tan\beta$, of the vacuum expectation values of two Higgs doublets.   Hence, there is a considerable
interest in studying the production mechanism of a single and a pair of Higgs bosons through bottom
quark annihilation.   The production of Higgs boson(s) in this mode is often studied using two approaches
namely four flavor and five flavor schemes often called 4FS and 5FS respectively.  In 4FS,
the bottom quark distribution in the proton is set 
equal to zero, however they are radiatively generated through 
gluons in the proton allowing the possibility of their annihilation to produce the Higgs boson.  
Such contributions are enhanced by large logarithms that are proportional to bottom quark mass and hence they
need to be resummed to get reliable predictions.  The alternate approach, 5FS, avoids this enhancement through
the introduction of non-zero bottom quark distributions in the proton.  The origin of these
distributions can be traced back to the bottom quarks resulting from gluon distributions inside the
proton, thanks to the DGLAP evolution equation of parton distribution functions, which resums collinear enhanced
logarithms to all orders in perturbation theory.  While both these schemes should give the
same result at the observable level, care is needed while comparing their predictions. 
The leading order contribution in 4FS comes from two to three scattering reaction
$g+g \rightarrow b +\overline b + H$ and the corresponding cross-section is of order $\alpha_s^2 \lambda^2$.
Here, the strong coupling constant $\alpha_s$ is given by $\alpha_s=g_s^2/4 \pi$ and 
the Yukawa coupling by $\lambda=m_b/v$ where $m_b$ is the mass of the bottom quark and 
$v$, the vacuum expectation value.
On the other hand, in 5FS, the leading order reaction is $b +\overline b\rightarrow H$ and 
the corresponding cross-section is proportional to only
$\lambda^2$ and is two orders less in the strong coupling constant. 
Hence, only at NNLO, the  
partonic cross-sections in the 5FS will have same order coupling strength, ${\cal O}(\alpha_s^2 \lambda^2)$,
as compared to the leading order (LO) cross-section in the 4FS. 
Because of this, the higher order computation in perturbative QCD 
in the 5FS is relatively easier.
Note that only next to leading order QCD effects \cite{Dittmaier:2003ej, Dawson:2003kb, Wiesemann:2014ioa}
are known in 4FS while in 5FS, 
the state of the art N$^3$LO prediction \cite{Duhr:2019kwi} is already available.  The later computation
provides an opportunity to compare N$^3$LO predictions against those at next-to-leading order (NLO) computed in 4FS in a
consistent manner.   In 5FS, the complete NLO \cite{Dicus:1998hs,Balazs:1998sb} and 
NNLO \cite{Harlander:2003ai} as well as dominant threshold effects at N$^3$LO 
\cite{Ravindran:2006cg,Ahmed:2014cha} are known for quiet sometime. 
In this article we improve the 5FS cross-section by resumming the threshold logarithms up to N$^3$LL accuracy.
It was observed in \cite{Bonvini:2016fgf,Forte:2015hba} that the 5FS cross-section provides 
dominant cross-section in a  matched prediction.   Thus threshold improved result justifies in the context of precision study for this process. Recently the 5FS prediction has been also improved \cite{Ebert:2017uel} by 
resumming time-like logarithms in SCET framework.

Fixed order (FO) predictions in perturbative QCD are often 
plagued with large logarithms resulting from certain boundaries
of the phase space and hence their reliability in those regions are questionable.
In the inclusive production rates, when partonic scaling variable $z = m_h^2/\hat s \rightarrow 1$, 
which corresponds to the emission of soft gluons, large logarithms are generated at every order in 
perturbation theory.  Here, $m_h$ is the Higgs boson mass and $\hat s$ is the partonic centre-of-mass
energy.  One finds a similar problem in the transverse momentum and rapidity distributions of
Higgs boson when there are soft gluon emissions. 
This is resolved by resumming these large logarithms to all orders in perturbation theory through
a systematic resummation approach.
For inclusive rates,  we refer \cite{Sterman:1986aj,Catani:1989ne,Catani:1990rp} to the earliest approach. 
Catani and Trentadue, in their seminal work \cite{Catani:1989ne},
demonstrated the resummation of leading large logarithms for the inclusive rates in
Mellin space.  

Needless to say that the inclusion of higher order QCD effects is of utmost 
importance to achieve precise prediction. 
In addition, these terms reduce the dependence on the unphysical scales such as 
renormalization and factorization scales at the observable level.
Note that for the production of scalar Higgs boson through gluon fusion,
the N$^3$LO contribution is now known~\cite{Anastasiou:2015ema,Anastasiou:2016cez},
which was further improved by the resummation of threshold logarithms,
arising from soft gluon emissions, to N$^3$LL$^\prime$ accuracy~\cite{Bonvini:2014joa,Bonvini:2016frm,Bonvini:2014tea}
(the prime $^\prime$ denotes that in addition to the N$^3$LL terms, the resummed result includes
higher logarithmic order terms coming from the matching to N$^3$LO).
Such an analysis leads to a precise determination of the SM Higgs cross-section at the LHC
with small uncertainty.

The goal of this paper is to present the prediction for the 
inclusive production of Higgs boson in bottom quark annihilation at the LHC taking into account the 
resummed threshold corrections at next-to-next-to-next to leading logarithmic accuracy.  We can achieve this
using the recent N$^3$LO prediction \cite{Duhr:2019kwi} and 
$N$ independent threshold constant $g_{b,0}$ computed to third order in QCD 
in this paper.  Here $N$ denotes the Mellin moment.  The $g_{b,0}^{(3)}$ is obtained using the
three loop form factor \cite{Gehrmann:2014vha} of the Higgs-bottom-anti-bottom operator and the 
third order soft distribution function computed in \cite{Ahmed:2014cha}.   

\section{Theoretical Framework} \label{theo}
The Lagrangian that describes the Yukawa interaction of the SM Higgs boson with  bottom quarks is given by 
\begin{align}
{\cal L}_{\rm int}^{\rm (S)} = -\lambda~ \bar{\psi}_{b}(x)\psi_{b}(x)\phi(x)
\end{align}
where $\psi_{b}(x)$ is the bottom quark field and $\phi(x)$ is the SM Higgs field.
$\lambda$ is the Yukawa coupling given by $\lambda = \frac{m_b}{v}$.
Note that we will use non-zero mass of the bottom quark only in the Yukawa coupling, elsewhere 
it is treated as massless quarks i.e. we use the VFS scheme throughout our analysis.
The inclusive cross-section for the production of Higgs boson in proton collision 
is given by
\begin{align}
\label{incsig}
\sigma(\tau,m_h^2) = \tau \sigma_{ b \bar{b}}^{(0)}(\mu_r^2) \sum_{ab = q,\overline q,g} \int {dx_1 \over x_1}
\int {dx_2\over x_2} f_a(x_1,\mu_f^2) f_b(x_2,\mu_f^2) \Delta_{ab}\left({\tau \over x_1 x_2},\mu_r^2,\mu_f^2\right)
\end{align}
where $f_c(x,\mu_f^2)$ is the non-perturbative parton distribution function (pdf) with $c$ denoting
the parton type and $x$ its momentum fraction.  
$\tau = m_h^2/S$ is the scaling variable, where $S$ is hadronic centre-of-mass energy. The renormalization and factorization scales are denoted by  $\mu_r$ and $\mu_f$ respectively.
The born cross-section $\sigma_{b \bar{b}}^{(0)}(\mu_r^2)$ is given by
\begin{align}
\sigma_{b\bar{b}}^{(0)}(\mu_r^2) = \frac{\pi m_b^2(\mu_r^2)}{6 m_h^2 v^2} \,.
\end{align}
The  mass factorized parton level cross-section ($\Delta_{ab}$) is calculable order by order in strong coupling constant, 
$a_s(\mu_r^2) = g_s^2(\mu_r^2)/16 \pi^2$ in perturbative QCD,
\begin{eqnarray}
\Delta_{ab}(z,\mu_r^2,\mu_f^2) = \delta_{a \bar{b}}\delta(1-z) + \sum_{i=1}^\infty a_s^i(\mu_r^2) 
\Delta^{(i)}_{ab}(z,\mu_r^2,\mu_f^2)\,.
\end{eqnarray}
At each order in perturbation theory one can write
\begin{eqnarray}
\Delta_{ab}^{(i)}(z,\mu_r^2,\mu_f^2) = \delta_{a\bar{b}} \Delta^{SV,(i)}(z,\mu_r^2,\mu_f^2) 
+ \Delta^{reg,{(i)}}_{ab}(z,\mu_r^2,\mu_f^2)
\end{eqnarray}
with $z = m_h^2/ \hat s$. 
In the above equation $\Delta^{SV}$ collects all those contributions that result from
soft and collinear configurations of partons in the scattering events.  They are 
proportional to distributions of the kind $\delta(1-z)$ and ${\cal D}_j(z)$ where
\begin{eqnarray}
{\cal D}_j(z) =\left( {\log^j(1-z) \over 1-z}\right)_+\,, \quad \quad \quad j=0,1,2,\cdots
\end{eqnarray}
The superscript $SV$  is the short form of soft plus virtual. 
The remaining contribution
is called the regular part of the cross-section denoted by $\Delta^{reg}_{ab}$.  
In QCD, $\Delta_{ab}$ was computed up to NNLO level in \cite{Harlander:2003ai}, 
at N$^3$LO level in the threshold limit
it was obtained in \cite{Ravindran:2006cg,Ahmed:2014cha} and recently the complete N$^3$LO result has been reported in \cite{Duhr:2019kwi}.  

The threshold contributions to inclusive cross-section $\Delta_{ab}(z)$ originate from soft and collinear
partons in the virtual and real emission subprocesses.   These contributions demonstrate remarkable
factorization property through process independent 
cusp, soft  and collinear anomalous dimensions.  Consequently, the leading contributions resulting
from the large threshold logarithms of the form ${\cal D}_j(z)$ can be summed to
all orders in a systematic fashion.   Following \cite{Sterman:1986aj,Catani:1989ne,Catani:1990rp}, 
the resummation of these logarithms can be
efficiently achieved in Mellin $N$-space and the resulting resummed threshold contribution 
takes the following form:   
\begin{eqnarray}
\label{delres}
\Delta^{res}_N(\mu_f^2)&=&\int_0^1 dz \ z^{N-1} \Delta^{res}(z,\mu_f^2)
\nonumber\\ 
&=& \tilde{g}_{b,0}(a_s(\mu_f^2)) \exp \left( 
\int_0^1 dz {z^{N-1} -1 \over 1-z} G(z,\mu_f^2)\right) 
\end{eqnarray}
where
\begin{eqnarray}
G(z,\mu_f^2) &=& \int_{\mu_f^2}^{q^2 (1-z)^2} {d \lambda^2 \over \lambda^2}
2 A^{(b)}(a_s(\lambda^2)) + D^{(b)}\left(a_s(q^2 (1-z)^2)\right)  
\end{eqnarray}
where we have set $\mu_r=\mu_f$.  The cusp anomalous dimension $A^{(b)}(a_s)$ and the constant $D^{(b)}(a_s)$
are process independent and hence can be obtained from the resummation result of Drell-Yan process. 
They are known up to third order in QCD and are listed in the Appendix \ref{apendc}.  
The $N$ independent constant $\tilde{g}_{b,0}(a_s(\mu_f^2))$
on the other hand is process dependent.  It gets contribution from the process dependent virtual 
Higgs-bottom-anti-bottom quark form factor.
The pre-factor $\tilde{g}_{b,0}(a_s(\mu_f^2))$, however gets modified due to the finite delta contribution from the soft function in Eq. (\ref{delres}) after integration. This leads to a new pre-factor $g_{b,0}(a_s(\mu_f^2))$. 
Following the method described in \cite{Ravindran:2006cg} and using 
the three loop form factor for the Higgs-bottom-anti-bottom quark 
\cite{Gehrmann:2014vha} and the third order soft distribution function \cite{Ahmed:2014cla}, 
we determine $g_{b,0}(a_s)$ up to third order in strong coupling constant. 
Expanding $g_{b,0}(a_s(\mu_f^2))$ in powers of $a_s(\mu_r^2)$ as
\begin{eqnarray}
g_{b,0}(a_s(\mu_f^2)) = 1+\sum_{i=1}^\infty a_s^i(\mu_r^2) g_{b,0}^{(i)}(\mu_r^2,\mu_f^2)\,,
\end{eqnarray}
we obtain $g_{b,0}^{(i)}$ for $i=1,2,3$:  

\begin{align} 
\begin{autobreak} 
\g01bbH = 
  \Cf    \bigg\{ 16  \z2
+ \bigg(-6\bigg)  \Lfr
- 4 \bigg\} ,   
\end{autobreak} 
\\ 
\begin{autobreak} 
\g02bbH = 
  \Ca  \Cf    \bigg\{ 
- \frac{92}{5}  \z2^2
+ \frac{256}{3}  \z2
+ \frac{280}{9}  \z3
+ \bigg(
- \frac{88}{3}  \z2
- 24  \z3
- 12\bigg)  \Lqr
+ \bigg(
- \frac{88}{3}  \z2
+ 24  \z3
- \frac{17}{3}\bigg)  \Lfr
+ \bigg(11\bigg)  \Lfr^2
+ \frac{166}{9} \bigg\}      
+ \Cf^2    \bigg\{ \frac{552}{5}  \z2^2
- 32  \z2
- 60  \z3
+ \bigg(
- 72  \z2
- 48  \z3
+ 21\bigg)  \Lfr
+ \bigg(
- 24  \z2
+ 48  \z3\bigg)  \Lqr
+ \bigg(18\bigg)  \Lfr^2
+ 16 \bigg\}      
+ \Cf  \nf    \bigg\{ 
- \frac{40}{3}  \z2
+ \frac{8}{9}  \z3
+ \bigg(\frac{16}{3}  \z2\bigg)  \Lqr
+ \bigg(\frac{16}{3}  \z2
+ \frac{2}{3}\bigg)  \Lfr
+ \bigg(-2\bigg)  \Lfr^2
+ \frac{8}{9} \bigg\} ,   
\end{autobreak} 
\\ 
\begin{autobreak} 
\g03bbH = 
  \Ca^2  \Cf    \bigg\{ \frac{7088}{63}  \z2^3
- \frac{25328}{135}  \z2^2
- \frac{7768}{9}  \z2  \z3
+ \frac{39980}{81}   \z2
- \frac{400}{3}  \z3^2
+ \frac{42748}{81}  \z3
- 84  \z5
+ \bigg(4  \z2^2
- \frac{8992}{27}  \z2
+ \frac{3104}{9}  \z3
- 80  \z5
+ \frac{1657}{18}\bigg)  \Lfr
+ \bigg(\frac{1964}{15}  \z2^2
- \frac{12800}{27}  \z2
- \frac{15472}{27}   \z3
+ 80  \z5
- \frac{1180}{3}\bigg)  \Lqr
+ \bigg(\frac{968}{9}  \z2
- 88  \z3
+ \frac{493}{9}\bigg)  \Lfr^2
+ \bigg(\frac{968}{9}   \z2
+ 88  \z3
+ 44\bigg)  \Lqr^2
+ \bigg(
- \frac{242}{9}\bigg)  \Lfr^3
+ \frac{68990}{81} \bigg\}      
+ \Ca  \Cf^2    \bigg\{ 
- \frac{123632}{315}  \z2^3
+ \frac{25676}{27}  \z2^2
+ \frac{2528}{3}  \z2  \z3
+ \frac{19658}{27}  \z2
+ \frac{592}{3}  \z3^2
- \frac{11188}{27}  \z3
- \frac{3352}{9}  \z5
+ \bigg(
- 472  \z2^2
- 352  \z2   \z3
- \frac{1748}{3}  \z2
+ \frac{3296}{3}  \z3
+ 240  \z5
+ \frac{388}{3}\bigg)  \Lqr
+ \bigg(
- \frac{1136}{5}  \z2^2
+ 352  \z2  \z3
- 212  \z2
- \frac{2536}{3}  \z3
- 240  \z5
- \frac{327}{2}\bigg)  \Lfr
+ \bigg(88  \z2
- 176   \z3\bigg)  \Lqr^2
+ \bigg(176  \z2
+ 144  \z3
+ 72\bigg)  \Lqrfr
+ \bigg(264  \z2
+ 32  \z3
+ 1\bigg)  \Lfr^2
+ \bigg(-66\bigg)  \Lfr^3
- \frac{982}{3} \bigg\}      
+ \Ca  \Cf  \nf    \bigg\{ \frac{184}{135}  \z2^2
+ \frac{880}{9}  \z2  \z3
- \frac{13040}{81}  \z2
- \frac{15944}{81}  \z3
- 8  \z5
+ \bigg(
- \frac{344}{15}  \z2^2
+ \frac{4480}{27}  \z2
+ \frac{3440}{27}  \z3
+ \frac{196}{3}\bigg)  \Lqr
+ \bigg(
- \frac{8}{5}  \z2^2
+ \frac{2672}{27}  \z2
- \frac{400}{9}  \z3
- 40\bigg)  \Lfr
+ \bigg(
- \frac{352}{9}  \z2
- 16  \z3
- 8\bigg)  \Lqr^2
+ \bigg(
- \frac{352}{9}  \z2
+ 16  \z3
- \frac{146}{9}\bigg)  \Lfr^2
+ \bigg(\frac{88}{9}\bigg)  \Lfr^3
- \frac{11540}{81} \bigg\}      
+ \Cf^3    \bigg\{ \frac{169504}{315}  \z2^3
- \frac{744}{5}  \z2^2
- 544  \z2  \z3
- \frac{166}{3}  \z2
+ 32   \z3^2
- 1188  \z3
+ 848  \z5
+ \bigg(
- \frac{1968}{5}  \z2^2
- 704  \z2  \z3
+ 12  \z2
+ 416   \z3
+ 480  \z5
- 113\bigg)  \Lfr
+ \bigg(
- \frac{1344}{5}  \z2^2
+ 704  \z2  \z3
+ 132  \z2
- 56  \z3
- 480  \z5
- 100\bigg)  \Lqr
+ \bigg(144  \z2
- 288  \z3\bigg)  \Lqrfr
+ \bigg(144  \z2
+ 288  \z3
- 54\bigg)  \Lfr^2
+ \bigg(-36\bigg)  \Lfr^3
+ \frac{1078}{3} \bigg\}      
+ \Cf^2  \nf    \bigg\{ 
- \frac{15688}{135}  \z2^2
- \frac{256}{3}  \z2  \z3
- \frac{3428}{27}  \z2
+ \frac{8872}{27}  \z3
- \frac{608}{9}  \z5
+ \bigg(\frac{272}{5}  \z2^2
+ 56  \z2
+ \frac{256}{3}  \z3
+ 38\bigg)  \Lfr
+ \bigg(\frac{464}{5}  \z2^2
+ \frac{200}{3}  \z2
- \frac{656}{3}  \z3
+ \frac{8}{3}\bigg)  \Lqr
+ \bigg(
- 48  \z2
- 32  \z3
- 4\bigg)  \Lfr^2
+ \bigg(
- 32  \z2\bigg)  \Lqrfr
+ \bigg(
- 16  \z2
+ 32  \z3\bigg)  \Lqr^2
+ \bigg(12\bigg)  \Lfr^3
- \frac{70}{9} \bigg\}      
+ \Cf  \nf^2    \bigg\{ \frac{448}{135}  \z2^2
+ \frac{256}{27}  \z2
+ \frac{160}{81}  \z3
+ \bigg(
- \frac{320}{27}  \z2
- \frac{64}{27}  \z3\bigg)  \Lqr
+ \bigg(
- \frac{160}{27}  \z2
+ \frac{32}{9}  \z3
+ \frac{34}{9}\bigg)  \Lfr
+ \bigg(\frac{32}{9}  \z2\bigg)  \Lqr^2
+ \bigg(\frac{32}{9}  \z2
+ \frac{4}{9}\bigg)  \Lfr^2
+ \bigg(
- \frac{8}{9}\bigg)  \Lfr^3
+ \frac{16}{27} \bigg\} , 
\end{autobreak} 
\end{align}
where $C_A = N$ and $C_F = (N^2-1)/2 N$ are the Casimirs of $SU(N)$ and 
$L_{fr} = \log(\mu_f^2/\mu_r^2)$ and $L_{qr} = \log(q^2 / \mu_r^2)$  with $q=m_h$.  
In addition $\zeta_i$ are the Riemann zeta functions.  
These constants $g_{b,0}^{(i)}$
can also be obtained from the soft-virtual results $\Delta^{SV,(i)}_{b \overline b}$ computed  
in \cite{Ahmed:2014cha} after taking the Mellin moment in 
$N$ space ( collected in App.\ \ref{App:B} ) and setting all the logarithms 
$\log(N)$ and Euler Gamma $\gamma_E$ to zero. 

To compute $G_N$, we use the method described in \cite{Catani:2003zt} up to N$^3$LL accuracy.   
Defining $\lambda=2 \beta_0 a_s(\mu_r^2) \ln(\overline N)$, we obtain 
\begin{eqnarray}\label{Eq:resumco1}
G_N &=& \int_0^1 dz {z^{N-1}-1 \over 1-z} G(z,\mu_f^2) \Bigg|_{N-{\rm dep}} 
\nonumber\\
&=&g_{1}(\lambda,\mu_r^2,\mu_f^2) \ln \bar{N} + \sum_{i=0}^\infty a_s^i(\mu_r^2) 
g_{i+2}(\lambda,\mu_r^2,\mu_f^2)\,,
\end{eqnarray}
where $\overline N = N \exp(\gamma_E)$ with $\gamma_E = 0.5772156649 \cdots$,
the Euler$-$Mascheroni constant. Note that $G$ contains only distributions in the exponent of Eq. 2.7 after taking out the delta contribution from the exponent.
The successive terms in the above equation defines the resummation accuracy LL, NLL etc. 
Note that in the context of resummation,  $\lambda$ is of ${\cal O}(1)$. The resummation coefficients in Eq.\ (\ref{Eq:resumco1}) agree with \cite{Catani:2003zt,Moch:2005ba} and for completeness we collect them 
in App.\ \ref{App:A}.

Thus, the resummed contribution to the Higgs production in bottom quark annihilation in Mellin $N$-space 
takes the simple form 
\begin{eqnarray}
\Delta^{res}_{N}(\mu_r^2,\mu_f^2) 
&=& g_{b,0}(\mu_r^2,\mu_f^2) \exp \left(G_N(\lambda,\mu_r^2,\mu_f^2)\right)\,.
\end{eqnarray}
The coefficients $g_{i}$ for $i=1,2,3,4$ are listed in the App.\ \ref{App:A}.  Note that except $A^{(b)}_4$
both the cusp $A^{(b)}$ and collinear $D^{(b)}$ anomalous dimensions are known to third order in $a_s$.
Since the approximate $A^{(b)}_4$ is available in the
literature \cite{Lee:2016ixa,Moch:2017uml,Grozin:2018vdn,Henn:2019rmi,Davies:2016jie,Lee:2017mip,Gracey:1994nn,Beneke:1995pq,Moch:2018wjh,Lee:2019zop} we can readily predict N$^3$LL contributions to inclusive Higgs production in bottom 
quark annihilation.  In the next section, using the recently available predictions \cite{Duhr:2019kwi}
for the fixed order N$^3$LO contributions, we present the resummed prediction to N$^3$LO + N$^3$LL accuracy.


\section{Phenomenology}
\label{nume}
In this section, we present a detailed discussion on the numerical impact 
of the resummed threshold contribution to the 
inclusive production of the Higgs boson in bottom quark annihilation at the
LHC up to N$^3$LO+N$^3$LL accuracy in perturbative QCD.     
The resummed part of the inclusive cross-section for the Higgs production in 
$N$ space can be obtained by taking $N$-th Mellin moment of Eq.~(\ref{incsig}) as
\begin{eqnarray}
\sigma^{res}_{N-1} = \sigma^{(0)}_{b\overline b}(\mu_r^2) \sum_{a=b,\overline b}
f_{a,N}(\mu_f^2) f_{\overline a,N}(\mu_f^2)
\Delta^{res}_N(\mu_r^2,\mu_f^2)  
\end{eqnarray}
where 
\begin{align}
\sigma^{res}_{N-1} &= \int_0^1 d \tau~ \tau^{N-2} \sigma^{res}(\tau,m_h^2) \,, \nonumber\\ 
f_{a,N} &= \int_0^1 dz ~z^{N-1} f_a(z,\mu_f^2)\,.
\end{align}
 $\sigma^{res}(\tau,m_h^2)$ is the
part of the cross-section where $\Delta_{ab} (z,\mu_r^2,\mu_f^2)$ in Eq.~(\ref{incsig}) is 
replaced by $\Delta^{res}(z,\mu_r^2,\mu_f^2)$ whose $N$-th moment
$\Delta^{res}_N$ is given in Eq. (\ref{delres}). 
The resummed cross-section is then added to the fixed order one 
after subtracting the Mellin moment of $\Delta^{SV,(n)}(z)$ in the large $N$ limit.
This is done because they are already present in the 
fixed order result and hence this will avoid double counting.
This is achieved through a matching procedure
at every order.  Finally, the matched result takes the following form: 
\begin{eqnarray}
\label{matched}
\sigma^{N^nLO+N^nLL}(\tau,m_h^2) &=& \sigma^{N^nLO}(\tau,m_h^2) 
+ \sigma^{(0)}_{b\bar{b}}(\mu_r^2) 
\int_{c-i\infty}^{c+i\infty} \frac{dN}{2 \pi i} \bigg( \frac{m_h^2}{S}\bigg)^{-N+1}
f_{b,N}(\mu_f^2) f_{\bar{b},N}(\mu_f^2) 
\nonumber\\
&&\times \bigg[ \Delta^{res,N^nLL}_N(\mu_r^2,\mu_f^2)  
- \Delta^{res,N^nLL}_N(\mu_r^2,\mu_f^2)\Big|_{\text{tr}} \bigg] \,.
\end{eqnarray}
In the above equation the superscript N$^n$LL in $\Delta^{res}$ implies that
we retain up to $g_{b,0}^{(n)}$ terms in the $g_{b,0}$ and up to $g_{n+1}(\lambda)$  
in the exponent $G_N$ given in Eq. (\ref{Eq:resumco1}).  Similarly,
N$^n$LO implies that we retain the fixed order result up to order $a_s^n$.  
The subscript $tr$ in the last term of the above equation means truncation
of the series in $a_s$ to desired accuracy.

The fixed order analytical results \cite{Harlander:2003ai} up to NNLO order have been implemented in a fortran
code.   We have validated our predictions 
to a very good accuracy with the available public code SuShi \cite{Harlander:2012pb}.
For N$^3$LO+N$^3$LL analysis, we have used the numbers presented in Tab.\ II in the arXiv version of the paper \cite{Duhr:2019kwi}
for N$^3$LO.
We perform the inverse Mellin transformation of the resummed $N$-space result using an
in-house fortran code.  Minimal prescription \cite{Catani:1996yz} has been used to deal with the Landau
pole in the Mellin inversion routine.
Since we work in the 5FS, we take $n_f = 5$ throughout. We use the MMHT2014(68cl) parton distribution 
set \cite{Harland-Lang:2014zoa}
and renormalization group (RG) running  for $a_s$ at each order.
One could in principal use the strong coupling as provided through LHAPDF \cite{Buckley:2014ana} interface, 
which at the NNLO level  changes the cross-section by $ 0.09\%$.
The Yukawa coupling  is evolved using 4-loop RG with bottom mass $m_b(m_b)=4.3$ GeV. 
It is well known that the optimal choice for the central scale to study the Higgs production 
in bottom quark annihilation is $\mu_r=m_h$ and $\mu_f=m_h/4$.  
This choice mimics most of the higher order contributions. 
Hence, we have made this choice throughout. 
In addition, we have predictions for other central scale choice namely $\mu_r=\mu_f=m_h$. 

\begin{figure*}[ht]
\centering{
\includegraphics[width=2.95in, height=3in]{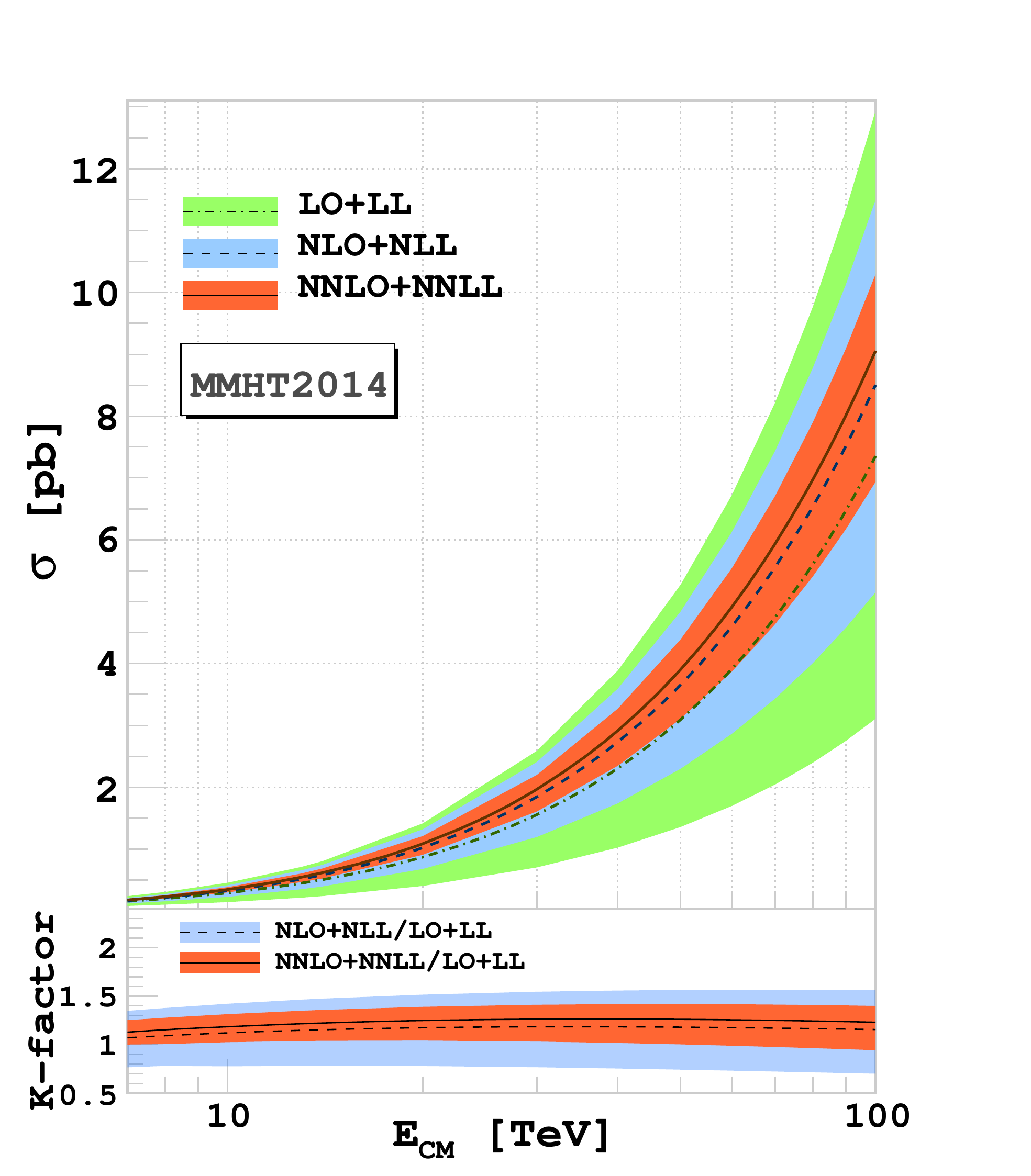}
\includegraphics[width=2.95in, height=3in]{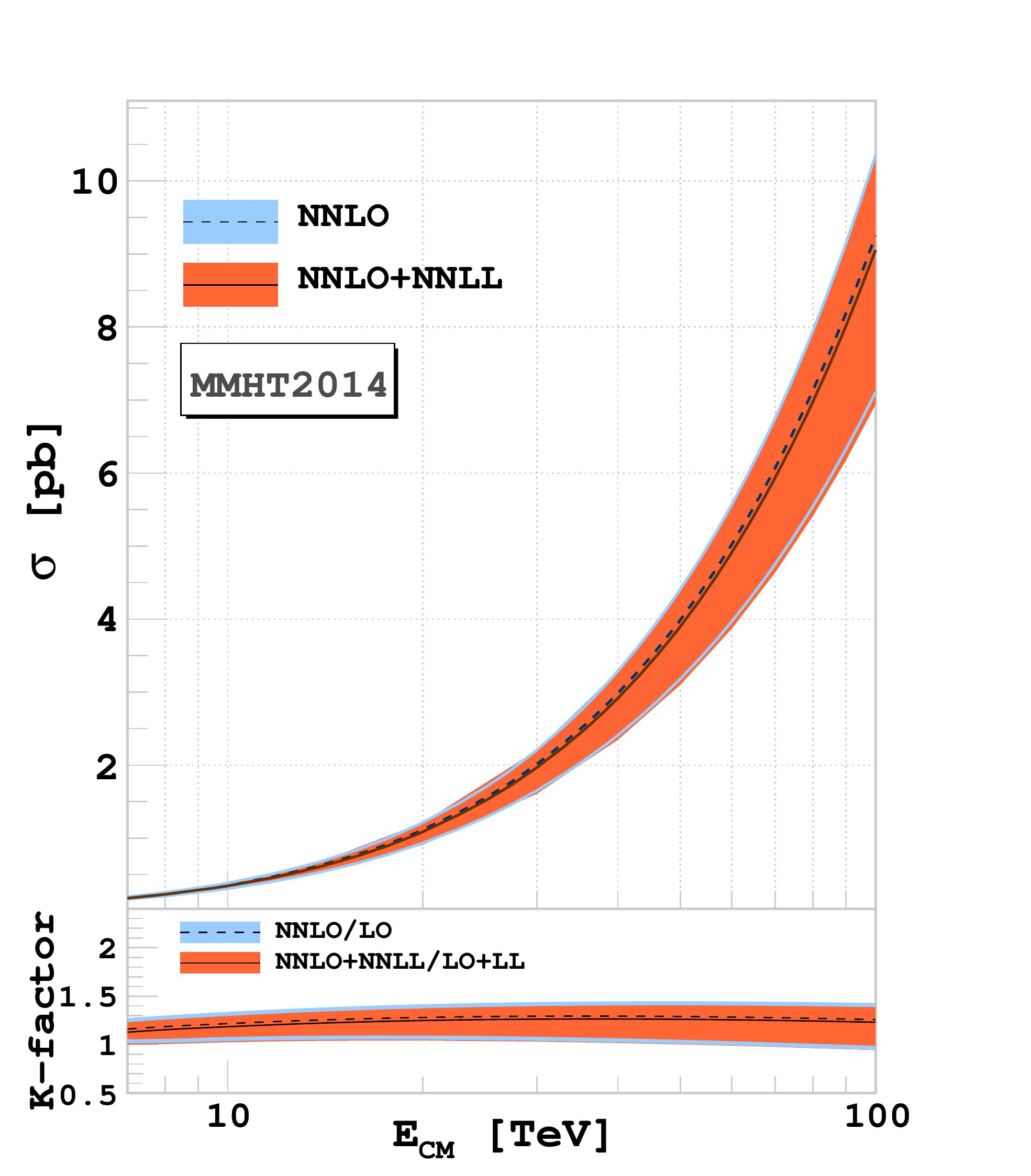}
}
\caption{Left: Resummed cross-section plotted against the hadronic center-of-mass energy ($E_{CM}$). The band corresponds to the scale variation around the central scale choice $(\mu_r^{(c)},\mu_f^{(c)}) = (1,1/4) m_h$ along with the prediction for the central scale. In the lower inset the resummed K-factor has been shown along with scale uncertainties (see text). Right: Same plot as the left but for the comparison of fixed order and resummed contribution at NNLO level.  }
\label{ecm_var}
\end{figure*}
In the left panel of Fig.\ \ref{ecm_var}, we present the resummed cross-section up to NNLO+NNLL level 
against the hadronic centre-of-mass energy for 7 TeV to 100 TeV.   
The bands in the plot correspond to the scale variations obtained by varying the 
unphysical renormalization and factorization scales in the range $(1/2,2) (\mu_r^{(c)},\mu_f^{(c)})$ 
where the central scales are taken to be $(\mu_r^{(c)},\mu_f^{(c)})= (1,1/4)~m_h$.  In the lower inset 
we show the corresponding resummed $k$-factors, defined as the ratio of the cross-section at a particular 
order (NLO+NLL, NNLO+NNLL) over the same at the LO+LL order. At NLL the $k$-factor increases by $14\%$ and at NNLL by $21\%$ compared to LL for $13$ TeV LHC.
We find that the uncertainty due to $\mu_r$ and $\mu_f$ scales increases with the 
energy of the collider, however at the current energy of the LHC 
it is within $11\%$ at NNLL.  The reason for the large uncertainty at high $E_{CM}$ could be due to the lack of knowledge of the pdf sets at these energies. In the right panel of Fig.\ \ref{ecm_var}, we compare the NNLO and NNLO+NNLL predictions for various centre-of-mass energies.  We find that the resummed cross-section
decreases by $3\%$ over the fixed order one for the entire range of centre-of-mass energies. 
The scale uncertainty however does not improve much compared to NNLO.

\begin{figure*}[ht]
\centering{
\includegraphics[width=5in, height=3in]{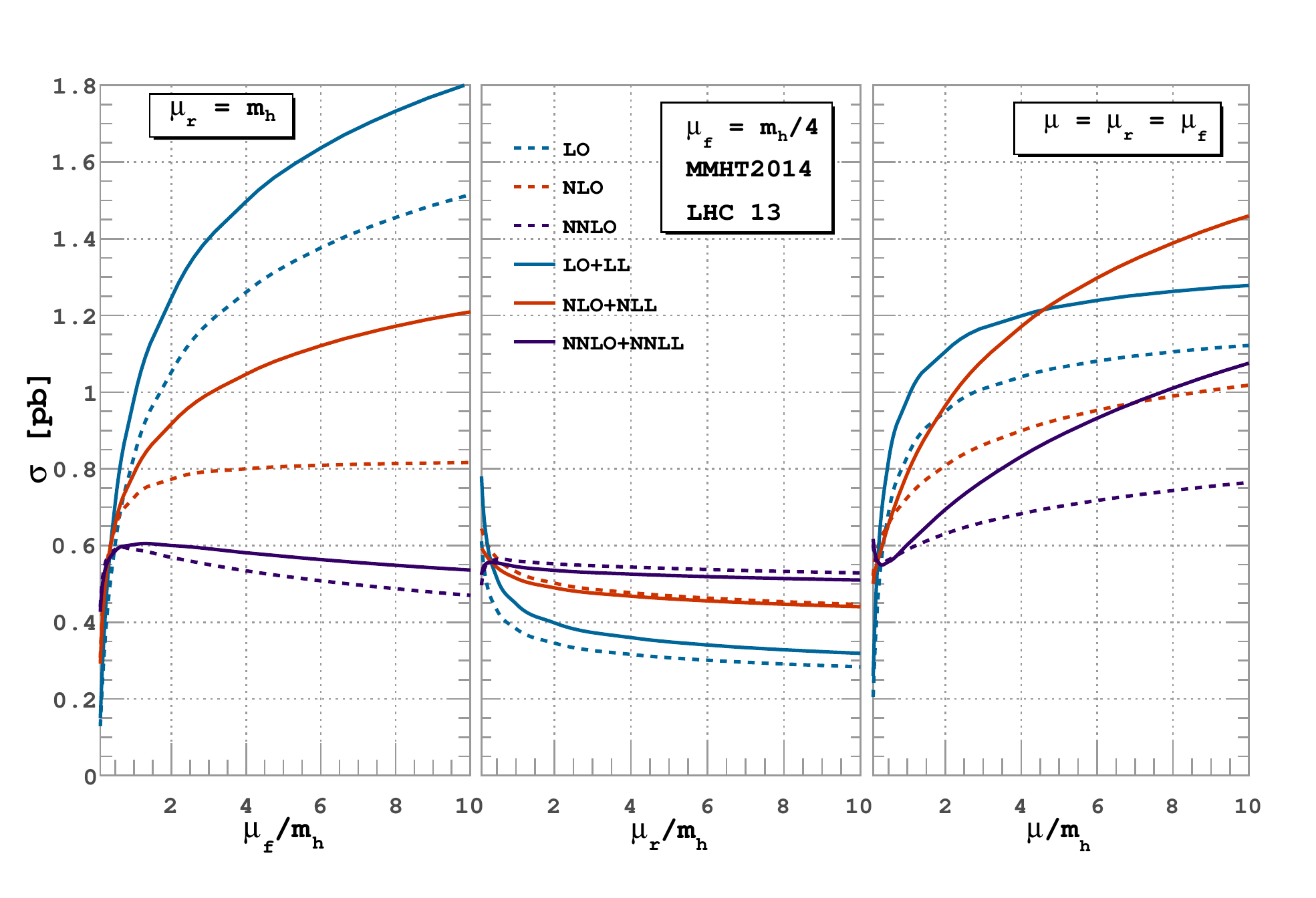}
}
\caption{Fixed order and resummed cross-sections are plotted successively at each order up to NNLO level with unphysical  factorization ($\mu_f$)   scale varied  in the range $(1/10,10) m_h$ keeping renormalization scale ($\mu_r$) fixed at central value $m_h$. Similar variation for $\mu_r$  is done in the second panel keeping $\mu_f=m_h/4$. In the last panel, the $\mu_r$ is set to $\mu_f$ and is varied in the same range.}
\label{mu_var}
\end{figure*}
In Fig.\ \ref{mu_var}, we present our findings on the sensitivity of our predictions 
to the renormalization and 
factorization scales at the LHC energy $13$ TeV. 
We plot the cross-sections at various orders as a function of $\mu_r$ and $\mu_f$ 
in the range $(0.1,10) m_h$. In the later part of our analysis 
we will choose the range $(1/2,2) m_h$.
In the first panel, we vary $\mu_f$ keeping $\mu_r=m_h$. 
We find that predictions for the cross-sections from fixed order as well as resummed ones are very 
sensitive to $\mu_f$ at first two orders and become better at third order. 
This could be due to lack of precise knowledge on the bottom quark distribution function.
At NNLO level, the uncertainty due to $\mu_f$ is about $^{-21}_{-16}\%$ in the 
range $\mu_f = (0.1,10)m_h$ and at NNLO+NNLL this goes down to $^{-21}_{-8}\%$. 
In the second panel we vary $\mu_r$ keeping the factorization scale $\mu_f$ at $m_h/4$. 
Interestingly, predictions from both the fixed order and the resummed one are 
less sensitive to $\mu_r$
compared to $\mu_f$.
In addition, as we increase the order of perturbation, the sensitivity to $\mu_r$ goes down significantly.
At LO, the cross-section decreases by $26\%$ when $\mu_r$ is varied ten times 
its central value. 
On the other hand it increases rapidly as we decrease $\mu_r$.  We find $59\%$ increase in the
cross-section when the scale is taken one tenth of the central value. 
In the resummed case, the corresponding changes are $^{-28\%}_{+67\%}$.
At NLO level the fixed order cross-section changes by $^{-16\%}_{+21\%}$ whereas for NLO+NLL case it improves and changes by $^{-15\%}_{+19\%}$.  
The corresponding numbers at NNLO are $^{-6\%}_{-11\%}$ for FO and  $^{-6.7\%}_{-4.9\%}$ for resummed one.
Finally in the third panel, we set $\mu_r=\mu_f=\mu$ and vary $\mu$ to see the combined effect.
We observe that the resultant uncertainty from the renormalization and factorization scales for 
resummed result compared to the one coming from fixed order result does not show reduction.  
This could be due to the fact that the resummation takes into account only the bottom quark
initiated channels beyond N$^3$LO accuracy and in addition the factorization scale dependence
dominates over the renormalisation scale.  The inclusion of the other channels 
and better determination of the bottom quark distribution can lead to the 
reduction of scale uncertainty.    

\begin{figure*}[ht]
\centering{
\includegraphics[width=5in, height=3in]{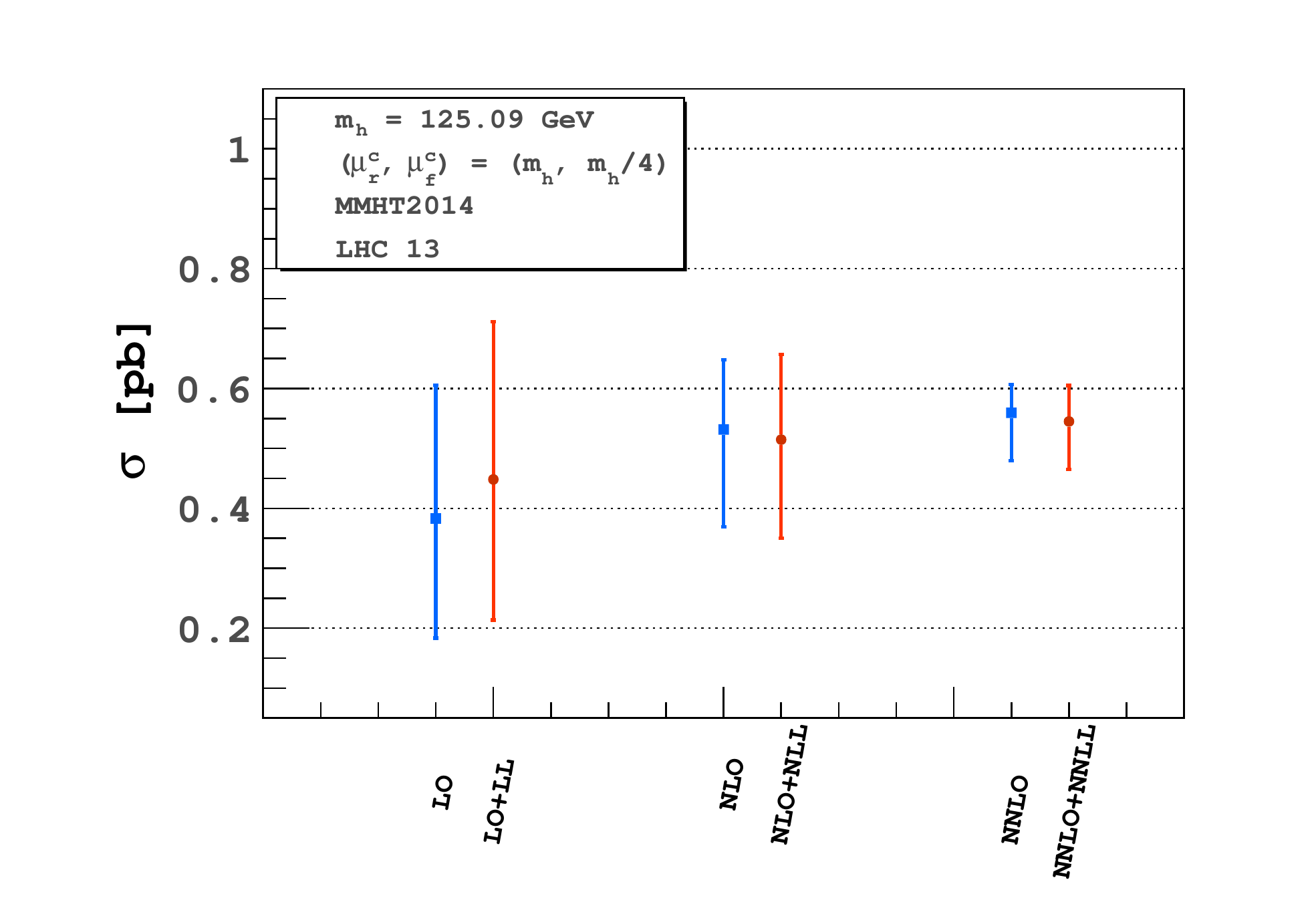}
}
\caption{The perturbative convergence is shown for fixed order and resummed order for $13$ TeV LHC. The central scale is fixed at $(1,1/4)m_h$ and the asymmetric error bars are obtained by varying the $\mu_r,\mu_f$ scales by $(1/2,2)$ around their central scale. }
\label{order}
\end{figure*}
In Fig.\ \ref{order}, we present the resummed cross-sections against the fixed
order ones for $13$ TeV LHC. The uncertainty at each order has been 
obtained by using the 7-point variation i.e., by varying $\mu_r$ and $\mu_f$ scales around 
their central values $m_h$ and $m_h/4$ respectively in the range $(1/2, 2)$. 
%
%
At NNLL level the uncertainty is comparable to the NNLO prediction.
However the central value of the resummed result shows a better perturbative convergence compared to the FO result.  
We find that the cross-section is larger by $38\%$ compared to that at LO 
whereas at NNLO, there is an increase of $5.2\%$ over the NLO prediction. 
In the resummed case, the NLO+NLL cross-section is $14.8\%$ larger than the LO+LL cross-section,  
while NNLO+NNLL is $5.8\%$ larger than NLO+NLL one.  
The perturbative convergence is seen to improve in the resummed case whereas the 
scale uncertainty does not improve compared to the fixed order.
This is mainly due to strong dependence of the cross-sections on the factorization scale at every
order in perturbation theory.  However, we expect that the uncertainty from $\mu_r$ will be mild. 

There are several pdfs that exist in the literature and the numerical predictions on the cross-section 
do depend on the choice of pdf.
A detailed analysis with respect to the choice of pdfs is thus required.
In particular, it will help us to understand how the underlying theoretical assumptions 
and models in the pdf fits affect the b-quark pdf determination \cite{Accardi:2016ndt}. 
We use some of the widely used pdf sets; e.g.\ the {\bf\texttt{ABMP16\_5\_nnlo}},  
{\bf \texttt{CT14nnlo}},  {\bf \texttt{MMHT2014nnlo68cl}},  {\bf \texttt{NNPDF31\_nnlo\_as\_0118}},  
{\bf \texttt{PDF4LHC15\_nnlo\_100}} pdf sets and corresponding strong coupling through LHAPDF interface.  
The central values for all pdf sets are given by the zeroth subset of the pdfs except from 
the NNPDF where the central value is given by the mean corresponding to all subsets. 
The results are tabulated in Table-\ref{tab:table1} with the central values and corresponding 
pdf uncertainties for different pdf groups.
We see that the central value as well as the uncertainties are small at NNLO+NNLL compared to the NNLO 
order for all the pdf sets.
The difference in the central prediction among different pdf groups can be 
traced back to the different treatment of bottom quark pdf by  different groups. 
We notice that the largest pdf uncertainty of $7.3\%$ comes from PDF4LHC sets whereas 
NNPDF sets give least uncertainty of $1.3\%$.  
We also observe that the pdf uncertainty is marginally improved in the case of NNLO+NNLL 
compared to that of NNLO. 
\begin{table}[h!]
\begin{center}
\begin{tabular}{|P{2.5cm}||P{3.5cm}|P{3.5cm}|} 
          \hline
      pdf sets & NNLO (pb) & NNLO+NNLL (pb)\\
     \hline \hline
       ABMP16        &   0.6265 $^{+0.0222}_{-0.0222}$  & 0.6113 $^{+0.0216}_{-0.0216}$ \\ \cline{1-3}
       CT14             &   0.5539 $^{+0.0256}_{-0.0274}$  & 0.5397 $^{+0.0249}_{-0.0267}$\\ \cline{1-3}
       MMHT2014   &   0.5581 $^{+0.0104}_{-0.0141}$  & 0.5437 $^{+0.0101}_{-0.0137}$\\ \cline{1-3}
       NNPDF31     &   0.5366  $^{+0.0070}_{-0.0070}$  & 0.5228 $^{+0.0068}_{-0.0068}$\\ \cline{1-3}
       PDF4LHC15 &   0.5863  $^{+0.0428}_{-0.0428}$   & 0.5712 $^{+0.0417}_{-0.0417}$\\
       \hline
\end{tabular}
\end{center}
\caption{Resummed cross-sections are provided for $13$ TeV LHC using NNLO pdf set for different pdf groups for central scale choice $(\mu_r^c,\mu_f^c) = (1,1/4) m_h$. The strong coupling is used from the respective pdfs through LHAPDF interface at NNLO. }
\label{tab:table1}
\end{table}



Recently, the fixed order predictions for $\Delta_{ab}$ to third order, namely $\Delta^{(3)}_{ab}$ 
have become available \cite{Duhr:2019kwi} and this allows us to predict the matched cross-section
to N$^3$LO+N$^3$LL accuracy using the approximate $A_4^{(b)}$. 
Note that SV contribution to N$^3$LO cross-section is already available, see \cite{Ahmed:2014cha}.
We have used same set of input parameters as in \cite{Duhr:2019kwi} for our study. 
For example, we use the PDF4LHC2015nnlo set with strong coupling evolved at four loops as in \cite{Duhr:2019kwi}. 
The bottom quark mass is taken to be $m_b(m_b)=4.18$ GeV. 
At the LHC with centre-of-mass energy $13$ TeV, we find that the threshold enhanced SV part of 
N$^3$LO cross-section differs 
from the complete N$^3$LO result by around $2\%$. 
Our prediction for the N$^3$LO$+$N$^3$LL cross-section at the central scale is $0.537$ pb 
which differs from the fixed order contribution by around $-1\%$.   
\begin{table}[h!]
\begin{center}
\begin{tabular}{|P{2.5cm}||P{3.5cm}|P{3.5cm}|} 
          \hline
      $E_{CM}$ (TeV)& NNLO+NNLL (pb) & N$^3$LO+N$^3$LL (pb)\\
     \hline \hline
        7  &   0.174  & 0.172\\ \cline{1-3}
        8  &   0.225  & 0.223\\ \cline{1-3}
       13 &  0.535  & 0.536\\ \cline{1-3}
       14 &  0.605  & 0.608\\
       \hline
\end{tabular}
\end{center}
\caption{Resummed cross-sections are provided for $7, 8, 13, 14$ TeV LHC with PDF4LHC2015nnlo pdf set for the central scale choice $(\mu_r^c,\mu_f^c) = (1,1/4) m_h$.}
\label{tab:table2}
\end{table}
In Table-\ref{tab:table2}, we quote the prediction at NNLO+NNLL and N$^3$LO+N$^3$LL levels 
for the central scale choice $(\mu_r^c,\mu_f^c) = (1,1/4)m_h$ at the centre-of-mass energies 
$7, 8, 13$ and $14$ TeV. We find that 
the central scale choice stabilizes irrespective of the collider energies already at NNLO+NNLL level. 
At N$^3$LO+N$^3$LL level, the change is well within $0.2\%$.

\begin{figure*}[ht]
\centering{
\includegraphics[width=5in, height=3in]{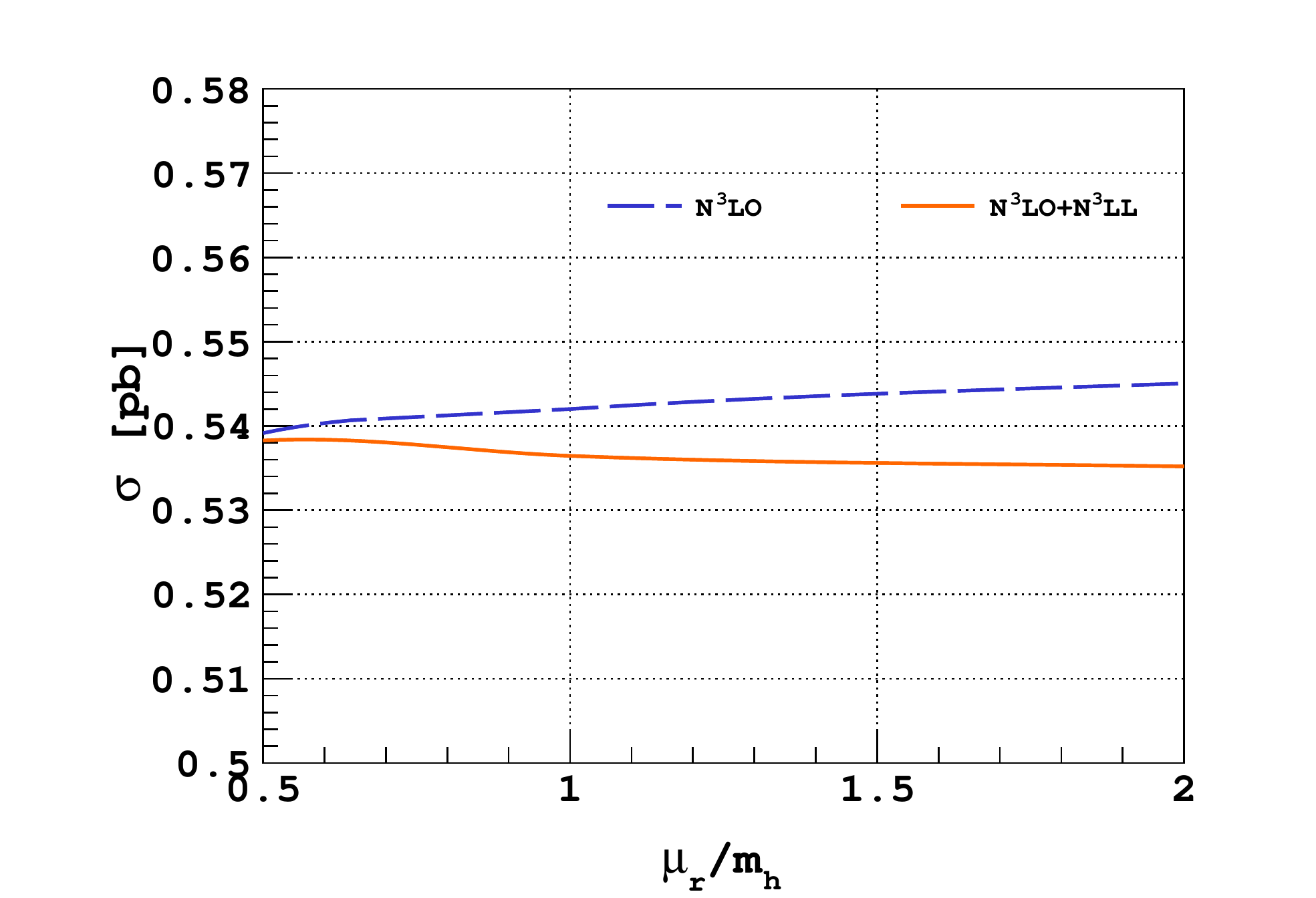}
}
\caption{Fixed order and resummed cross-sections are plotted at  N$^3$LO level against unphysical  renormalization ($\mu_r$)   scale varied  in the range $(1/2,2) m_h$ keeping factorization scale ($\mu_f$) fixed at central value $m_h/4$. All the parameters are chosen same as in \cite{Duhr:2019kwi}.}
\label{mu_var2}
\end{figure*}
A detailed study on the dependence of the cross-section at N$^3$LO$+$N$^3$LL level on the 
renormalization and factorization scales is possible provided the numerical 
code at N$^3$LO \cite{Duhr:2019kwi} is publicly available. 
In addition, for the $\mu_f$ variation, one also needs the pdfs at the N$^3$LO.
However, using the renormalization group equation 
and the numerical predictions 
given in table-II of \cite{Duhr:2019kwi} 
for a selected set of values of centre-of-mass energies, 
it is possible 
to study the dependence on renormalization scale.  
In Fig.\ \ref{mu_var2}, we show how 
predictions at N$^3$LO and N$^3$LO+N$^3$LL depend  
on the renormalization scale.  In fact, entire dependence from $\mu_r$ at N$^3$LO level is
controlled by the contributions from previous orders, namely LO, NLO and NNLO.
We observe that the sensitivity to renormalization scale goes down at N$^3$LO level 
compared to the previous order. At NNLO level the $\mu_r$ variation changes the cross-section 
by $^{+0.8}_{-1.3} \%$ compared to the central value for the variation 
in the range $\mu_r \in \{1/2,2\}m_h$. 
At N$^3$LO level this decreases to $^{-0.5}_{+0.6}\%$. 
For the resummed prediction, we find that  the dependence on $\mu_r$ goes further down at 
N$^3$LO+N$^3$LL level.  The corresponding uncertainty is $^{+0.3}_{-0.2} \%$ 
compared to the central value. This shows the improvement in the 
stability against $\mu_r$ at N$^3$LO+ N$^3$LL level compared to fixed order.

\section{Conclusions}\label{conc}
In this article we have studied the role of resummed threshold corrections to the inclusive 
cross-section for producing the Higgs boson through bottom quark annihilation at the LHC. 
While this is a sub-dominant channel compared to gluon fusion subprocess, the precise measurements 
that are carried out at the LHC in the context of processes involving Higgs bosons demand 
the inclusion of this channel for precision studies.  Complete NNLO QCD corrections \cite{Harlander:2003ai} 
and soft plus virtual corrections \cite{Ahmed:2014cha}
to N$^3$LO level to this observable are known for a while.    
More recently, a complete N$^3$LO contributions resulting
from all the partonic channels where the Higgs boson couples to bottom quarks became 
available \cite{Duhr:2019kwi}. 
In hadronic colliders, soft gluons play an important role in most of the observables. 
In the fixed order perturbative computations, in certain kinematic regions the soft gluons 
dominate.   The large logarithms resulting from these soft gluons often spoil 
the reliability of the fixed order predictions.  The resolution to this
problem is to resum these logarithms to all orders in perturbation theory.  The framework to resum 
such logarithms in a systematic fashion to all orders in perturbation theory for inclusive observables 
is well established and results based on this demonstrate better and reliable predictions compared  to the fixed order ones.  
In the present article we have done a detailed study to understand the role of the soft gluons 
on the inclusive cross-section for producing Higgs boson in bottom quark annihilation.  This is
achieved within the frame work of threshold resummation in Mellin-N space.  We have done this to N$^3$LO+N$^3$LL accuracy. 
We have used the recent prediction at N$^3$LO level from \cite{Duhr:2019kwi} for the fixed order
contribution and for the resummed part at N$^3$LL level, except the process dependent constant
$g_{b,0}^{(3)}$, rest of the ingredients are already known.  
We have computed this constant for the first time using the three loop form factor 
\cite{Gehrmann:2014vha} and the quark soft distribution function \cite{Ahmed:2014cha} known to third
order in QCD.  Our numerical result at N$^3$LO+N$^3$LL in QCD is the most precise prediction for 
the inclusive cross-section for the production of Higgs boson through bottom quarks at the LHC.   
We have predicted the inclusive rates at various centre-of-mass energies with the corresponding $k$-factors. 
In addition, we studied in detail the sensitivity of the resummed predictions 
to the renormalization and factorization scales and choice of parton distribution functions 
in order to estimate the theoretical errors precisely.

\section*{Acknowledgements}
We would like to thank P. K. Dhani for initial collaboration. We  also  thank S. Moch, F. J. Tackmann and J. K. L. Michel for useful discussions.  
\appendix
\section{Resummation constants $g_i(\lambda,\mu_r^2,\mu_f^2)$}\label{App:A}
The resumamtion constants $g_i$  (see \ref{Eq:resumco1}) in the $N$-space are found to be  
\begin{align} 
\begin{autobreak} 
\gNB1 =
\bigg[ \AAo  ~  \bigg\{ 2
- 2 ~ \LogmW1
+ 2 ~ \LogmW1 ~ \iW \bigg\}
\bigg],   
\end{autobreak} 
\\ 
\begin{autobreak} 
\gNB2 =
\bigg[ \DDo  ~  \bigg\{ \frac{1}{2} ~ \LogmW1 \bigg\}      
+ \AAt  ~  \bigg\{ 
- \LogmW1
- \w \bigg\}      
+ \AAo  ~  \bigg\{ \bigg(\LogmW1
+ \frac{1}{2} ~ \LogmW1^2
+ \w\bigg) ~ \bigg(\frac{\beta_{1}}{\beta_0^{2}}\bigg)
+ \bigg(\w\bigg) ~ \Lfr
+ \bigg(\LogmW1\bigg) ~ \Lqr \bigg\}
\bigg],   
\end{autobreak} 
\\ 
\begin{autobreak} 
\gNB3 =
\bigg[ \btzAIII  ~  \bigg\{ \frac{1}{2} ~ \WbimW
- \frac{1}{2} ~ \w \bigg\}      
+ \btzAII  ~  \bigg\{ \bigg(
- \frac{3}{2} ~ \WbimW
- \LogomWtIMW
+ \frac{1}{2} ~ \w\bigg) ~ \bigg(\frac{\beta_{1}}{\beta_0^{2}}\bigg)
+ \bigg(
- \WbimW\bigg) ~ \Lqr
+ \bigg( \w\bigg) ~ \Lfr \bigg\}      
+ \btzAI  ~  \bigg\{ 2 ~ \z2 ~ \WbimW
+ \bigg(\frac{1}{2} ~ \LogtmWtIMW
+ \frac{1}{2} ~ \WbimW
+ \LogomWtIMW
- \LogmW1
- \frac{1}{2} ~ \w\bigg) ~ \bigg(\frac{\beta_{1}}{\beta_0^{2}}\bigg)^2
+ \bigg(\frac{1}{2} ~ \WbimW\bigg) ~ \Lqr^2
+ \bigg(\frac{1}{2} ~ \WbimW
+ \LogmW1
+ \frac{1}{2} ~  \w\bigg) ~ \btoo
+ \bigg(
- \frac{1}{2} ~ \w\bigg) ~ \Lfr^2
+ \bigg(\bigg(\WbimW
+ \LogomWtIMW\bigg) ~ \bigg(\frac{\beta_{1}}{\beta_0^{2}}\bigg)\bigg) ~ \Lqr \bigg\}      
+ \btzDII  ~  \bigg\{ 
- \frac{1}{2} ~ \WbimW \bigg\}      
+ \btzDI  ~  \bigg\{ \bigg(\frac{1}{2} ~ \WbimW\bigg) ~ \Lqr
+ \bigg(\frac{1}{2} ~ \WbimW
+ \frac{1}{2} ~ \LogomWtIMW\bigg) ~ \bigg(\frac{\beta_{1}}{\beta_0^{2}}\bigg) \bigg\}
\bigg],   
\end{autobreak} 
\\ 
\begin{autobreak} 
\gNB4 =
\bigg[ \btztAIV  ~  \bigg\{ \frac{1}{6} ~ \WttmWtimWt
- \frac{1}{3} ~ \w \bigg\}      
+ \btztAIII  ~  \bigg\{ \bigg(
- \frac{1}{2} ~ \WttmWtimWt\bigg) ~ \Lqr
+ \bigg(
- \frac{5}{12} ~ \WttmWtimWt
- \frac{1}{2} ~  \LogomWtIMWt
+ \frac{1}{3} ~ \w\bigg) ~ \bigg(\frac{\beta_{1}}{\beta_0^{2}}\bigg)
+ \bigg(\w\bigg) ~ \Lfr \bigg\}      
+ \btztAII  ~  \bigg\{ 2 ~ \z2 ~ \WttmWtimWt
+ \bigg(\frac{1}{2} ~ \LogtmWtIMWt
- \frac{1}{12} ~ \WtbimWt
+ \frac{5}{6} ~  \WbimW
+ \frac{1}{2} ~ \LogomWtIMWt
- \frac{1}{3} ~ \w\bigg) ~ \bigg(\frac{\beta_{1}}{\beta_0^{2}}\bigg)^2
+ \bigg(\frac{1}{2} ~ \WttmWtimWt\bigg) ~ \Lqr^2
+ \bigg(\frac{1}{3}  ~ \WtbimWt
- \frac{1}{3} ~ \WbimW
+ \frac{1}{3} ~ \w\bigg) ~ \btoo
+ \bigg(
- \w\bigg) ~ \Lfr^2
+ \bigg(\bigg(\frac{1}{2} ~ \WttmWtimWt
+ \LogomWtIMWt\bigg) ~ \bigg(\frac{\beta_{1}}{\beta_0^{2}}\bigg)\bigg) ~ \Lqr \bigg\}      
+ \btztAI  ~  \bigg\{ \frac{8}{3} ~ \z3 ~ \WttmWtimWt
+ \bigg(
- \frac{1}{6} ~ \LogttmWtIMWt
+ \frac{1}{3} ~ \WtbimWt
- \frac{1}{3} ~ \WbimW
+ \frac{1}{2} ~ \LogomWtIMWt
- \LogomWtIMW
+ \frac{1}{2} ~ \LogmW1
+ \frac{1}{3} ~ \w\bigg) ~ \bigg(\frac{\beta_{1}}{\beta_0^{2}}\bigg)^3
+ \bigg(
- \frac{1}{6} ~ \WttmWtimWt\bigg) ~ \Lqr^3
+ \bigg(\frac{1}{12} ~ \WttmWtimWt
+ \frac{1}{2} ~ \LogmW1
+ \frac{1}{3} ~ \w\bigg) ~ \bthr
+ \bigg(
- \frac{5}{12} ~ \WtbimWt
+ \frac{1}{6} ~ \WbimW
- \frac{1}{2} ~ \LogomWtIMWt
+ \LogomWtIMW
- \LogmW1
- \frac{2}{3} ~ \w\bigg) ~ \bobt
+ \bigg(\frac{1}{3} ~ \w\bigg) ~ \Lfr^3
+ \bigg(
- 2 ~ \z2 ~ \WttmWtimWt
+ \bigg(
- \frac{1}{2} ~  \LogtmWtIMWt
+ \frac{1}{2} ~ \WtbimWt\bigg) ~ \bigg(\frac{\beta_{1}}{\beta_0^{2}}\bigg)^2
+ \bigg(
- \frac{1}{2} ~ \WtbimWt\bigg) ~ \btoo\bigg) ~ \Lqr
+ \bigg(
- 2 ~ \z2  ~ \LogomWtIMWt\bigg) ~ \bigg(\frac{\beta_{1}}{\beta_0^{2}}\bigg)
+ \bigg(\bigg(
- \frac{1}{2} ~ \LogomWtIMWt\bigg) ~ \bigg(\frac{\beta_{1}}{\beta_0^{2}}\bigg)\bigg) ~ \Lqr^2
+ \bigg(\bigg(
- \frac{1}{2} ~ \w\bigg) ~ \bigg(\frac{\beta_{1}}{\beta_0^{2}}\bigg)\bigg) ~  \Lfr^2 \bigg\}      
+ \btztDIII  ~  \bigg\{ 
- \frac{1}{4} ~ \WttmWtimWt \bigg\}      
+ \btztDII  ~  \bigg\{ \bigg(\frac{1}{4} ~ \WttmWtimWt
+ \frac{1}{2} ~ \LogomWtIMWt\bigg) ~ \bigg(\frac{\beta_{1}}{\beta_0^{2}}\bigg)
+ \bigg(\frac{1}{2} ~ \WttmWtimWt\bigg) ~  \Lqr \bigg\}      
+ \btztDI  ~  \bigg\{ 
- \z2 ~ \WttmWtimWt
+ \bigg(
- \frac{1}{4} ~ \LogtmWtIMWt
+ \frac{1}{4} ~ \WtbimWt\bigg) ~ \bigg(\frac{\beta_{1}}{\beta_0^{2}}\bigg)^2
+ \bigg(
- \frac{1}{4} ~ \WttmWtimWt\bigg) ~ \Lqr^2
+ \bigg(
- \frac{1}{4} ~ \WtbimWt\bigg) ~ \btoo
+ \bigg(\bigg(
- \frac{1}{2} ~  \LogomWtIMWt\bigg) ~ \bigg(\frac{\beta_{1}}{\beta_0^{2}}\bigg)\bigg) ~ \Lqr \bigg\}
\bigg]\,. 
\end{autobreak} 
\end{align}

\section{Soft-Virtual coefficients in $N$-space}\label{App:B}
Here we have collected all the large $N$ coefficients for this process. Defining $\bar{L}=\ln N + \gamma_E$ they are given below:

\begin{align} 
\begin{autobreak} 
\gSVN1 = 
  \LNb^2    \bigg\{ \bigg(8\bigg)  \Cf \bigg\}     
+ \LNb    \bigg\{ \bigg(\bigg(-8\bigg)  \Lqr
+ \bigg(8\bigg)  \Lfr\bigg)  \Cf \bigg\}      + g_{01}, 
\end{autobreak} 
\\ 
\begin{autobreak} 
\gSVN2 = 
  \LNb^4    \bigg\{ \bigg(32\bigg)  \Cf^2 \bigg\}      
+ \LNb^3    \bigg\{ \bigg(\bigg(-64\bigg)  \Lqr
+ \bigg(64\bigg)  \Lfr\bigg)  \Cf^2
+ \bigg(
- \frac{32}{9}\bigg)  \Cf  \nf
+ \bigg(\frac{176}{9}\bigg)  \Ca   \Cf \bigg\}      
+ \LNb^2    \bigg\{ \bigg(
- 16  \z2
+ \bigg(
- \frac{88}{3}\bigg)  \Lqr
+ \frac{536}{9}\bigg)  \Ca  \Cf
+ \bigg(128  \z2
+ \bigg(-64\bigg)   \Lqrfr
+ \bigg(-48\bigg)  \Lfr
+ \bigg(32\bigg)  \Lqr^2
+ \bigg(32\bigg)  \Lfr^2
- 32\bigg)  \Cf^2
+ \bigg(\bigg(\frac{16}{3}\bigg)  \Lqr
- \frac{80}{9}\bigg)  \Cf  \nf \bigg\}      
+ \LNb    \bigg\{ \bigg(
- 56  \z3
+ \bigg(
- 16  \z2
+ \frac{536}{9}\bigg)  \Lfr
+ \bigg(16  \z2
- \frac{536}{9}\bigg)  \Lqr
+ \bigg(
- \frac{44}{3}\bigg)  \Lfr^2
+ \bigg(\frac{44}{3}\bigg)  \Lqr^2
+ \frac{1616}{27}\bigg)  \Ca  \Cf
+ \bigg(\bigg(
- 128  \z2
+ 32\bigg)   \Lqr
+ \bigg(128  \z2
- 32\bigg)  \Lfr
+ \bigg(-48\bigg)  \Lfr^2
+ \bigg(48\bigg)  \Lqrfr\bigg)  \Cf^2
+ \bigg(\bigg(
- \frac{80}{9}\bigg)   \Lfr
+ \bigg(
- \frac{8}{3}\bigg)  \Lqr^2
+ \bigg(\frac{8}{3}\bigg)  \Lfr^2
+ \bigg(\frac{80}{9}\bigg)  \Lqr
- \frac{224}{27}\bigg)  \Cf  \nf \bigg\}      + g_{02},
\end{autobreak} 
\\ 
\begin{autobreak} 
\gSVN3 = 
  \LNb^6    \bigg\{ \bigg(\frac{256}{3}\bigg)  \Cf^3 \bigg\}      
+ \LNb^5    \bigg\{ \bigg(\bigg(-256\bigg)  \Lqr
+ \bigg(256\bigg)  \Lfr\bigg)  \Cf^3
+ \bigg(
- \frac{256}{9}\bigg)  \Cf^2  \nf
+ \bigg(\frac{1408}{9}\bigg)  \Ca  \Cf^2 \bigg\}      
+ \LNb^4    \bigg\{ \bigg(
- 128  \z2
+ \bigg(
- \frac{3520}{9}\bigg)  \Lqr
+ \bigg(\frac{1408}{9}\bigg)  \Lfr
+ \frac{4288}{9}\bigg)  \Ca   \Cf^2
+ \bigg(512  \z2
+ \bigg(-512\bigg)  \Lqrfr
+ \bigg(-192\bigg)  \Lfr
+ \bigg(256\bigg)  \Lqr^2
+ \bigg(256\bigg)   \Lfr^2
- 128\bigg)  \Cf^3
+ \bigg(\bigg(
- \frac{256}{9}\bigg)  \Lfr
+ \bigg(\frac{640}{9}\bigg)  \Lqr
- \frac{640}{9}\bigg)  \Cf^2  \nf
+ \bigg(
- \frac{704}{27}\bigg)  \Ca  \Cf  \nf
+ \bigg(\frac{64}{27}\bigg)  \Cf  \nf^2
+ \bigg(\frac{1936}{27}\bigg)  \Ca^2  \Cf \bigg\}      
+ \LNb^3    \bigg\{ \bigg(
- \frac{704}{9}  \z2
+ \bigg(
- \frac{3872}{27}\bigg)  \Lqr
+ \frac{28480}{81}\bigg)  \Ca^2  \Cf
+ \bigg(
- \frac{512}{9}  \z2
+ \bigg(
- \frac{1088}{9}\bigg)  \Lfr
+ \bigg(\frac{64}{3}\bigg)  \Lfr^2
+ \bigg(\frac{128}{3}\bigg)  \Lqrfr
+ \bigg(\frac{1280}{9}\bigg)   \Lqr
+ \bigg(-64\bigg)  \Lqr^2
- \frac{1696}{27}\bigg)  \Cf^2  \nf
+ \bigg(\frac{128}{9}  \z2
+ \bigg(\frac{1408}{27}\bigg)  \Lqr
- \frac{9248}{81}\bigg)  \Ca  \Cf  \nf
+ \bigg(\frac{2816}{9}  \z2
- 448  \z3
+ \bigg(
- 256  \z2
+ \frac{7520}{9}\bigg)  \Lfr
+ \bigg(256  \z2
- \frac{8576}{9}\bigg)  \Lqr
+ \bigg(
- \frac{704}{3}\bigg)  \Lqrfr
+ \bigg(
- \frac{352}{3}\bigg)  \Lfr^2
+ \bigg(352\bigg)   \Lqr^2
+ \frac{10816}{27}\bigg)  \Ca  \Cf^2
+ \bigg(\bigg(
- 1024  \z2
+ 256\bigg)  \Lqr
+ \bigg(1024  \z2
- 256\bigg)   \Lfr
+ \bigg(
- \frac{256}{3}\bigg)  \Lqr^3
+ \bigg(\frac{256}{3}\bigg)  \Lfr^3
+ \bigg(-384\bigg)  \Lfr^2
+ \bigg(-256\bigg)   \Lqrfrt
+ \bigg(256\bigg)  \Lqrtfr
+ \bigg(384\bigg)  \Lqrfr\bigg)  \Cf^3
+ \bigg(\bigg(
- \frac{128}{27}\bigg)  \Lqr
+ \frac{640}{81} \bigg)  \Cf  \nf^2 \bigg\}      
+ \LNb^2    \bigg\{ \bigg(
- \frac{2016}{5}  \z2^2
+ \frac{15296}{9}  \z2
+ \frac{2240}{9}  \z3
+ \bigg(
- 704  \z2
+ 256   \z3
- \frac{12352}{27}\bigg)  \Lqr
+ \bigg(
- \frac{416}{3}  \z2
- 256  \z3
+ \frac{2056}{27}\bigg)  \Lfr
+ \bigg(
- 128   \z2
+ \frac{4288}{9}\bigg)  \Lqr^2
+ \bigg(
- 128  \z2
+ \frac{5080}{9}\bigg)  \Lfr^2
+ \bigg(256  \z2
- \frac{6992}{9}\bigg)   \Lqrfr
+ \bigg(
- \frac{352}{3}\bigg)  \Lqr^3
+ \bigg(
- \frac{352}{3}\bigg)  \Lfr^3
+ \bigg(\frac{352}{3}\bigg)  \Lqrtfr
+ \bigg(\frac{352}{3} \bigg)  \Lqrfrt
- \frac{272}{3}\bigg)  \Ca  \Cf^2
+ \bigg(\frac{352}{5}  \z2^2
- \frac{2144}{9}  \z2
- 352  \z3
+ \bigg(\frac{352}{3}   \z2
- \frac{14240}{27}\bigg)  \Lqr
+ \bigg(\frac{968}{9}\bigg)  \Lqr^2
+ \frac{62012}{81}\bigg)  \Ca^2  \Cf
+ \bigg(\frac{4416}{5}   \z2^2
- 256  \z2
- 480  \z3
+ \bigg(
- 1024  \z2
+ 256\bigg)  \Lqrfr
+ \bigg(
- 576  \z2
- 384   \z3
+ 168\bigg)  \Lfr
+ \bigg(
- 192  \z2
+ 384  \z3\bigg)  \Lqr
+ \bigg(512  \z2
- 128\bigg)  \Lqr^2
+ \bigg( 512  \z2
+ 16\bigg)  \Lfr^2
+ \bigg(-192\bigg)  \Lfr^3
+ \bigg(-192\bigg)  \Lqrtfr
+ \bigg(384\bigg)  \Lqrfrt
+ 128\bigg)  \Cf^3
+ \bigg(
- \frac{2240}{9}  \z2
+ \frac{640}{9}  \z3
+ \bigg(\frac{128}{3}  \z2
- \frac{208}{27}\bigg)  \Lfr
+ \bigg(128   \z2
+ \frac{1648}{27}\bigg)  \Lqr
+ \bigg(
- \frac{784}{9}\bigg)  \Lfr^2
+ \bigg(
- \frac{640}{9}\bigg)  \Lqr^2
+ \bigg(
- \frac{64}{3}\bigg)   \Lqrtfr
+ \bigg(
- \frac{64}{3}\bigg)  \Lqrfrt
+ \bigg(\frac{64}{3}\bigg)  \Lqr^3
+ \bigg(\frac{64}{3}\bigg)  \Lfr^3
+ \bigg(\frac{992}{9}\bigg)   \Lqrfr
- \frac{92}{3}\bigg)  \Cf^2  \nf
+ \bigg(\frac{320}{9}  \z2
+ \bigg(
- \frac{64}{3}  \z2
+ \frac{4624}{27}\bigg)  \Lqr
+ \bigg(
- \frac{352}{9}\bigg)  \Lqr^2
- \frac{16408}{81}\bigg)  \Ca  \Cf  \nf
+ \bigg(\bigg(
- \frac{320}{27}\bigg)  \Lqr
+ \bigg(\frac{32}{9}\bigg)  \Lqr^2
+ \frac{800}{81}\bigg)  \Cf  \nf^2 \bigg\}      
+ \LNb    \bigg\{ \bigg(
- \frac{176}{5}  \z2^2
+ \frac{352}{3}  \z2  \z3
- \frac{12784}{81}  \z2
- \frac{24656}{27}  \z3
+ 384  \z5
+ \bigg(
- \frac{352}{5}  \z2^2
+ \frac{2144}{9}  \z2
+ 352  \z3
- \frac{62012}{81}\bigg)  \Lqr
+ \bigg(\frac{352}{5}   \z2^2
- \frac{2144}{9}  \z2
+ \frac{176}{3}  \z3
+ \frac{980}{3}\bigg)  \Lfr
+ \bigg(
- \frac{176}{3}  \z2
+ \frac{7120}{27}\bigg)   \Lqr^2
+ \bigg(\frac{176}{3}  \z2
- \frac{7120}{27}\bigg)  \Lfr^2
+ \bigg(
- \frac{968}{27}\bigg)  \Lqr^3
+ \bigg(\frac{968}{27}\bigg)   \Lfr^3
+ \frac{594058}{729}\bigg)  \Ca^2  \Cf
+ \bigg(
- \frac{32}{5}  \z2^2
+ \frac{1648}{81}  \z2
+ \frac{1808}{27}  \z3
+ \bigg(
- \frac{320}{9}  \z2
+ \frac{16408}{81}\bigg)  \Lqr
+ \bigg(
- \frac{32}{3}  \z2
+ \frac{2312}{27}\bigg)  \Lfr^2
+ \bigg(\frac{32}{3}  \z2
- \frac{2312}{27}\bigg)  \Lqr^2
+ \bigg(\frac{320}{9}  \z2
- \frac{224}{3}  \z3
- \frac{1672}{27}\bigg)  \Lfr
+ \bigg(
- \frac{352}{27}\bigg)  \Lfr^3
+ \bigg(\frac{352}{27}\bigg)  \Lqr^3
- \frac{125252}{729}\bigg)  \Ca  \Cf  \nf
+ \bigg(\frac{64}{5}  \z2^2
- \frac{3584}{27}  \z2
+ \frac{608}{9}  \z3
+ \bigg(
- \frac{2240}{9}  \z2
+ \frac{640}{9}  \z3
+ \frac{172}{9}\bigg)  \Lfr
+ \bigg(
- \frac{256}{3}   \z2
+ \frac{8}{3}\bigg)  \Lqr^2
+ \bigg(\frac{256}{3}  \z2
+ 56\bigg)  \Lfr^2
+ \bigg(\frac{2240}{9}  \z2
- \frac{640}{9}  \z3
+ \frac{92}{3}\bigg)  \Lqr
+ \bigg(
- \frac{176}{3}\bigg)  \Lqrfr
+ \bigg(-32\bigg)  \Lfr^3
+ \bigg(16\bigg)  \Lqrtfr
+ \bigg(16\bigg)  \Lqrfrt
- \frac{842}{9}\bigg)  \Cf^2  \nf
+ \bigg(
- 896  \z2  \z3
+ \frac{25856}{27}  \z2
+ 224  \z3
+ \bigg(
- \frac{2016}{5}   \z2^2
+ \frac{15296}{9}  \z2
+ \frac{5264}{9}  \z3
- \frac{4048}{9}\bigg)  \Lfr
+ \bigg(\frac{2016}{5}  \z2^2
- \frac{15296}{9}   \z2
- \frac{2240}{9}  \z3
+ \frac{272}{3}\bigg)  \Lqr
+ \bigg(
- \frac{1120}{3}  \z2
+ 192  \z3
- 344\bigg)  \Lfr^2
+ \bigg(
- 96  \z2
- 384  \z3
+ \frac{920}{3}\bigg)  \Lqrfr
+ \bigg(\frac{1408}{3}  \z2
+ 192  \z3
+ \frac{112}{3}\bigg)   \Lqr^2
+ \bigg(-88\bigg)  \Lqrtfr
+ \bigg(-88\bigg)  \Lqrfrt
+ \bigg(176\bigg)  \Lfr^3
- \frac{6464}{27}\bigg)  \Ca  \Cf^2
+ \bigg(\frac{64}{9}  \z3
+ \bigg(
- \frac{800}{81}\bigg)  \Lqr
+ \bigg(
- \frac{160}{27}\bigg)  \Lfr^2
+ \bigg(
- \frac{32}{27}\bigg)  \Lqr^3
+ \bigg(
- \frac{32}{27}\bigg)  \Lfr
+ \bigg(\frac{32}{27}\bigg)  \Lfr^3
+ \bigg(\frac{160}{27}\bigg)  \Lqr^2
+ \frac{3712}{729}\bigg)  \Cf   \nf^2
+ \bigg(\bigg(
- \frac{4416}{5}  \z2^2
+ 256  \z2
+ 480  \z3
- 128\bigg)  \Lqr
+ \bigg(\frac{4416}{5}  \z2^2
- 256  \z2
- 480  \z3
+ 128\bigg)  \Lfr
+ \bigg(
- 576  \z2
- 384  \z3
+ 168\bigg)  \Lfr^2
+ \bigg( 192  \z2
- 384  \z3\bigg)  \Lqr^2
+ \bigg(384  \z2
+ 768  \z3
- 168\bigg)  \Lqrfr
+ \bigg(-144\bigg)   \Lqrfrt
+ \bigg(144\bigg)  \Lfr^3\bigg)  \Cf^3 \bigg\}    + g_{03},  
\end{autobreak} 
\end{align}

\section{The Cusp and the soft anomalous dimensions}\label{App:C}
The quark cusp anomalous dimensions are given as \cite{Moch:2004pa},
\label{apendc}

\begin{align} 
\begin{autobreak} 
A_1^{(b)} = C_{F}
\bigg\{ 4
\bigg\},   
\end{autobreak} 
\\ 
\begin{autobreak} 
A_2^{(b)} = C_{F}
\bigg\{ \nf    \bigg( 
- \frac{40}{9} \bigg)      
+ \Ca    \bigg( \frac{268}{9}
- 8  \z2 \bigg)
\bigg\},   
\end{autobreak} 
\\ 
\begin{autobreak} 
A_3^{(b)} = C_{F}
\bigg\{ \nf^2    \bigg( 
- \frac{16}{27} \bigg)      
+ \Cf  \nf    \bigg( 
- \frac{110}{3}
+ 32  \z3 \bigg)      
+ \Ca  \nf    \bigg( 
- \frac{836}{27}
- \frac{112}{3}  \z3
+ \frac{160}{9}  \z2 \bigg)      
+ \Ca^2    \bigg( \frac{490}{3}
+ \frac{88}{3}  \z3
- \frac{1072}{9}  \z2
+ \frac{176}{5}  \z2^2 \bigg)
\bigg\}\,.
\end{autobreak} 
\end{align}

The four-loop coefficient \cite{Moch:2018wjh,Lee:2017mip,Grozin:2018vdn,Henn:2019rmi,Lee:2019zop,vonManteuffel:2019wbj} is also known numerically and the perturbative series for $A^{(b)}$ finally looks like
 \begin{align}
  A^{(b)}(n_f\!=\!3) &=
  0.42441\:\alpha_s \:
  ( 1  +  0.7266\,\alpha_s +  0.7341\,\alpha_s^2 + 0.665\,\alpha_s^3
        + \, \dots )
\; , \nonumber \\
  A^{(b)}(n_f\!=\!4) &=
  0.42441\:\alpha_s \:
  ( 1  +  0.6382\,\alpha_s  +  0.5100\,\alpha_s^2 + 0.317\,\alpha_s^3
       + \, \dots )
\; , \nonumber \\
  A^{(b)}(n_f\!=\!5) &=
  0.42441\:\alpha_s \:
  ( 1  +  0.5497\,\alpha_s  +  0.2840\, \alpha_s^2 + 0.013\,\alpha_s^3
        + \, \dots )
\, . \nonumber \\
\end{align}
The resummed coefficients $D_i^{(b)}$ are given as,

\begin{align} 
\begin{autobreak} 
D_1^{(b)} = {C}_{F}
\bigg\{0
\bigg\},   
\end{autobreak} 
\\ 
\begin{autobreak} 
D_2^{(b)} = {C}_{F}
\bigg\{ \nf    \bigg( \frac{224}{27}
- \frac{32}{3}  \z2 \bigg)      
+ \Ca    \bigg( 
- \frac{1616}{27}
+ 56  \z3
+ \frac{176}{3}  \z2 \bigg)
\bigg\},   
\end{autobreak} 
\\ 
\begin{autobreak} 
D_3^{(b)} = { C}_{F}
\bigg\{ \nf^2    \bigg( 
- \frac{3712}{729}
+ \frac{320}{27}  \z3
+ \frac{640}{27}  \z2 \bigg)      
+ \Cf  \nf    \bigg( \frac{3422}{27}
- \frac{608}{9}  \z3
- 32  \z2
- \frac{64}{5}  \z2^2 \bigg)      
+ \Ca  \nf    \bigg( \frac{125252}{729}
- \frac{2480}{9}  \z3
- \frac{29392}{81}  \z2
+ \frac{736}{15}  \z2^2 \bigg)      
+ \Ca^2    \bigg( 
- \frac{594058}{729}
- 384  \z5
+ \frac{40144}{27}  \z3
+ \frac{98224}{81}  \z2
- \frac{352}{3}   \z2  \z3
- \frac{2992}{15}  \z2^2 \bigg)
\bigg\}\,.
\end{autobreak} 
\end{align}

\bibliographystyle{JHEP}
\bibliography{bbHresv2}
\end{document}